\documentclass[12pt,letterpaper,makeidx]{article}
\usepackage[latin1]{inputenc}
\usepackage{amsmath}
\usepackage{amsfonts}
\usepackage{amssymb}
\usepackage{epsfig}
\usepackage{theorem}
\usepackage{oldgerm}
\usepackage{tabularx}
\usepackage{placeins}
\usepackage{amscd}
\usepackage{bm}
\usepackage{epic}
\usepackage{curves}
\usepackage{graphicx}
\usepackage{makeidx}
\usepackage{fancyhdr}
\usepackage{graphpap}
\usepackage{rotating}
\usepackage{bbm}
\usepackage{pstricks}
\usepackage[absolute]{textpos}
\usepackage{psfrag}
\usepackage{times}
\usepackage{dcolumn}
\usepackage{shadow}
\usepackage{multicol}
\usepackage[latin1]{inputenc}
\usepackage{oldgerm}
\fancypagestyle{plain}{%
\fancyhead{} 
}
\title{Charge conjugation from space-time inversion in QED: discrete and continuous groups}  
\author{B. Carballo Pérez\footnote{brendacp@nucleares.unam.mx}  and M. Socolovsky\footnote{socolovs@nucleares.unam.mx} \\ 
\small{ Instituto de Ciencias Nucleares, Universidad Nacional Autónoma de México,}\\
\small{ Circuito exterior, Ciudad Universitaria, 04510, México D.F., México}}
\date{}

\begin{document}
\maketitle
\textbf{Abstract}\\
\hspace{0.5cm} We show that the CPT groups of QED emerge naturally from the $\cal{PT}$ and $\cal{P}$ (or $\cal{T}$) subgroups of the Lorentz group. We also find relationships between these discrete groups and continuous groups, like the connected Lorentz and Poincaré groups and their universal coverings.\\


\newpage

\section{Introduction}

\hspace{0.5cm} It was shown in \cite{CPT group} that the CPT group, $G_{\hat{\theta}}(\hat{\psi})$ ($\hat{\theta}=\hat{C}*\hat{P}*\hat{T}$), of the Dirac quantum field is a non abelian group with sixteen elements isomorphic to the direct product of the quaternion group, $Q$, and the cyclic group, $\mathbb{Z}_2$:
\begin{equation}
G_{\hat{\theta}}(\hat{\psi})\cong Q \times \mathbb{Z}_2.
\end{equation} 

Unlike $G_{\hat{\theta}}(\hat{\psi})$ \cite{CPT group, B2, B1}, the CPT group, $G_{\hat{\theta}}(\hat{A_{\mu}})$, of the electromagnetic field is an abelian group of eight elements with three generators \cite{B2}:
\begin{equation}
G_{\hat{\theta}}(\hat{A_{\mu}})\cong\mathbb{Z}_2^{3}.
\end{equation}

As the CPT transformation properties of the interacting $\hat{\psi}-\hat{A_{\mu}}$ fields are the same as for the free fields \cite{Azcarraga}, the complete CPT group for QED, $G_{\hat{\Theta}}(QED)$, is the direct product of the above mentioned two groups, $G_{\hat{\Theta}}(\hat{\psi})$ and $G_{\hat{\Theta}}(\hat{A_{\mu}})$, i.e.,

\begin{equation}
G_{\hat{\Theta}}(QED)=G_{\hat{\Theta}}(\hat{\psi})\times G_{\hat{\Theta}}(\hat{A_{\mu}})\cong (Q \times \mathbb{Z}_{2})\times \mathbb{Z}_{2}^{3}.
\end{equation}

\section{C from $\mathbf{\cal{PT}}$}
\hspace{0.5cm} It was shown in \cite{B1} that $Q$ becomes isomorphic to a subgroup $H$ of $SU(2)$, being $\lambda$ the isomorphism:
\begin{eqnarray}
Q \buildrel {\lambda}\over \longrightarrow H < SU(2),\nonumber\\
1 \mapsto I, \quad \iota \mapsto -i \sigma_{1},  \quad \gamma\mapsto -i \sigma_{2},  \quad  \kappa\mapsto -i \sigma_{3},
\label{lambda}
\end{eqnarray}
where $\iota$, $\gamma$, $\kappa$ are the three imaginary units of the quaternion group and $\sigma_k$ ($k=1,2,3$) are the Pauli matrices; and taking also into account that $\mathbb{Z}_2$ is isomorphic to the center of $SU(2)$: $\lbrace I,-I \rbrace$, then:

\begin{equation}
G_{\hat{\theta}}(\hat{\psi}) \cong H\times (center \ \ of \ \ SU(2)).
\end{equation}

Since $SU(2)$ is the universal covering group of  $SO(3)$:

\begin{equation}
SU(2) \buildrel {\Phi}\over \longrightarrow SO(3),
\label{Phi}
\end{equation}
then $\Phi(H)$ has 4 elements and, for that reason, the unique candidates are groups isomorphic to $C_4$ and $D_2\cong \mathbb{Z}_2\times \mathbb{Z}_2$, the Klein group. A simple application of $\Phi$ to the elements of $H$ led to: 
\begin{equation}
\Phi (H)=\{I, \ R_x(\pi), \ R_y(\pi), \ R_z(\pi)\},
\end{equation}
with $R_x(\pi)$, $R_y(\pi)$, $R_z(\pi)$ the rotations in $\pi$ around the axes $x$, $y$ and $z$, respectively, and $I$, the unit matrix in $SO(3)$. It was then immediately verified that the multiplication table of $\Phi(H) < SO(3)$ is the same as for $D_2$.

Then, we have:
\begin{equation}
G_{\hat{\theta}}(\hat{\psi}) \cong \Phi^{-1}(D_2)\times \mathbb{Z}_2. 
\end{equation} 

Within the Lorentz group $O(3,1)$, the transformations of parity $\cal{P}$ and time reversal $\cal{T}$, together with their product $\cal{PT}$ and the 4$\times$4 unit matrix $E$, lead to the subgroup of the Lorentz group, called the $\cal{PT}$-group, which is also isomorphic to $D_2$.
 
On the other hand, $\cal{P}$ or $\cal{T}$ separately, together with the unit 4$\times$4 matrix $E$, give rise to the group $\mathbb{Z}_{2}$. Then, we obtain the desired result for the Dirac field: 

\begin{equation}
G_{\hat{\theta}}(\hat{\psi})\cong \Phi^{-1}(<\lbrace {\cal{P}},{\cal{T}} \rbrace>)\times <\lbrace {\cal{P}} \rbrace> 
\end{equation} 
or
\begin{equation}
G_{\hat{\theta}}(\hat{\psi})\cong \Phi^{-1}(<\lbrace {\cal{P}},{\cal{T}} \rbrace>)\times <\lbrace {\cal{T}} \rbrace>;
\end{equation} 
while, for the electromagnetic field, we have:
\begin{equation}
G_{\hat{\Theta}}(\hat{A_{\mu}})\cong <\lbrace {\cal{P}},{\cal{T}} \rbrace>\times<\lbrace {\cal{P}} \rbrace>
\end{equation}
or
\begin{equation}
G_{\hat{\Theta}}(\hat{A_{\mu}})\cong <\lbrace {\cal{P}},{\cal{T}} \rbrace>\times<\lbrace {\cal{T}} \rbrace>.
\end{equation}

The above result suggests that the Minkowskian space-time structure of special relativity, in particular the unconnected component of its symmetry group, the real Lorentz group $O(3,1)$, implies the existence of the CPT group as a whole, and therefore the existence of the charge conjugation transformation, and thus the proper existence of antiparticles.

\section{Discrete and continuous groups}
\hspace{0.5cm}The relationships between the discrete groups: $Q$, $G_{PT}=<\lbrace {\cal{P}},{\cal{T}} \rbrace>$, $G_{\hat{\theta}}(\hat{\psi})$ and $G_{\hat{\Theta}}(\hat{A_{\mu}})$ and continuous groups, like the Lorentz group and its universal covering group, can be summarized in the following diagram:
\begin{eqnarray}
\begin{array}[c]{cccccccccc}   
    \mathbb{Z}_{2} &  & \mathbb{Z}_{2} & & \mathbb{Z}_{2} &  & \mathbb{Z}_{2} & & \mathbb{Z}_{2}\\
    \downarrow\scriptstyle{}& &\downarrow\scriptstyle{} & &\downarrow\scriptstyle{} & &\downarrow\scriptstyle{} & &\downarrow\scriptstyle{} & \\
    G_{\hat{\theta}}(\hat{\psi})\cong Q \times \mathbb{Z}_2 & \stackrel{\alpha}{\leftarrow} & Q & \stackrel{\mu}{\rightarrow} & SU(2) & \stackrel{\beta}{\rightarrow} & SL_{2}(\mathbb{C}) & \stackrel{\gamma}{\rightarrow} & \mathbb{R}^{4}\rtimes SL_{2}(\mathbb{C})\\
    \downarrow\scriptstyle{\psi}& &\downarrow\scriptstyle{\rho} & &\downarrow\scriptstyle{\Phi} & &\downarrow\scriptstyle{\overset{\sim}{\Phi}} & &\downarrow\scriptstyle{\overset{\approx}{\Phi}} &\\
    G_{\hat{\theta}}(\hat{A_{\mu}})\cong\mathbb{Z}_2^{3} & \stackrel{\bar{\alpha}}{\leftarrow} & G_{PT}\cong \mathbb{Z}_2^{2}& \stackrel{\bar{\mu}}{\rightarrow} & SO(3) & \stackrel{\bar{\beta}}{\rightarrow} & SO^{c}(3,1) & \stackrel{\bar{\gamma}}{\rightarrow} & \mathbb{R}^{4}\rtimes SO^{c}(3,1).
\end{array}
\label{diagrama1}
\end{eqnarray}

The homomorphism $\mu$ is defined by $\mu(q)=\lambda(q)$ (see (\ref{lambda})) and the homomorphism $\Phi$ was described in (\ref{Phi}); $\overset{\sim}{\Phi}$ and $\overset{\approx}{\Phi}$ are the homomorphisms between the conected Lorentz ($SO^{c}(3,1)$) and Poincaré ($\mathbb{R}^{4}\rtimes SO^{c}(3,1) \equiv {\cal{P}}_{4}^{c}$) groups, respectively, and their universal coverings ($SL_{2}(\mathbb{C})$ and $\mathbb{R}^{4}\rtimes SL_{2}(\mathbb{C})\equiv \bar{\cal{P}}_{4}^{c}$); while $\rho$, $\psi$, $\bar{\mu}$, $\alpha$, $\bar{\alpha}$, $\beta$, $\bar{\beta}$, $\gamma$ and $\bar{\gamma}$ are given by:

\begin{eqnarray}
Q \buildrel {\rho}\over \longrightarrow \frac{Q}{\mathbb{Z}_{2}}\cong  G_{PT}, \quad q \mapsto [q],
\label{rho}
\end{eqnarray}
\begin{eqnarray}
G_{\hat{\theta}}(\hat{\psi})\buildrel {\psi}\over \longrightarrow \frac{Q\times\mathbb{Z}_{2}}{\mathbb{Z}_{2}}\cong G_{\hat{\theta}}(\hat{A_{\mu}}), \quad (q,1) \mapsto [(q,1)], \quad (q,-1) \mapsto [(q,1)],
\label{psi}
\end{eqnarray}
\begin{eqnarray}
G_{PT} \buildrel {\bar{\mu}}\over \longrightarrow  SO(3), \quad [q] \mapsto \Phi(h),\quad h=\lambda(q),
\label{mu}
\end{eqnarray}
\begin{eqnarray}
Q \buildrel {\alpha}\over \longrightarrow G_{\hat{\theta}}(\hat{\psi}), \quad q \mapsto (q,1),
\label{alpha}
\end{eqnarray}
\begin{eqnarray}
G_{PT} \buildrel {\bar{\alpha}}\over \longrightarrow G_{\hat{\theta}}(\hat{A_{\mu}}), \quad [q] \mapsto [(q,1)],
\label{baralpha}
\end{eqnarray}
\begin{eqnarray}
SU(2) \buildrel {\beta}\over \longrightarrow SL_{2}(\mathbb{C}), \quad A \mapsto A,
\label{beta}
\end{eqnarray}
\begin{eqnarray}
SO(3) \buildrel {\bar{\beta}}\over \longrightarrow SO^{c}(3,1), \quad R \mapsto 
\begin{pmatrix}
1 & 0 \cr 0 & R \cr
\end{pmatrix},
\label{barbeta}
\end{eqnarray}
\begin{eqnarray}
SL_{2}(\mathbb{C}) \buildrel {\gamma}\over \longrightarrow \mathbb{R}^{4}\rtimes SL_{2}(\mathbb{C}), \quad B \mapsto (\bold{0},B),
\label{gamma}
\end{eqnarray}
\begin{eqnarray}
SO^{c}(3,1) \buildrel {\bar{\gamma}}\over \longrightarrow \mathbb{R}^{4}\rtimes SO^{c}(3,1), \quad \Lambda \mapsto (\bold{0},\Lambda).
\label{bargamma}
\end{eqnarray}

Let $\nu$ the function which goes from $Q\times \mathbb{Z}_{2}$ to $SU(2)$:
\begin{eqnarray}
Q\times \mathbb{Z}_{2} \buildrel {\nu}\over \longrightarrow SU(2), \quad (q,g) \mapsto \nu(q,g):=sg(g)\lambda(q),
\label{gamma}
\end{eqnarray}
where $sg(g)=1$ if $g=1$ and $sg(g)=-1$ if $g=-1$.

Then, it holds:
\begin{itemize}
\item $\nu$ is an homomorphism.\\
Proof:
\begin{eqnarray}
\nu((q',g')(q,g))&=&\nu(q'q,g'g)=sg(g'g)\lambda(q'q)=sg(g')sg(g)\lambda(q')\lambda(q)\nonumber\\
&=&(sg(g')\lambda(q'))(sg(g)\lambda(q))=\nu(q',g')\nu(q,g).
\end{eqnarray}
\item $\nu$ is 2 to 1.\\
Proof:
\begin{equation}
\nu(q,-1)=\nu(-q,1).
\end{equation}
\end{itemize}

$\bar{\nu}$ is determined by $\nu$  due to the commutative diagram:
\begin{eqnarray}
\begin{array}[c]{cccccc}   
    G_{\hat{\theta}}(\hat{\psi})  & \stackrel{\nu}{\rightarrow} & SU(2)\\
    \downarrow\scriptstyle{\psi}&  &\downarrow\scriptstyle{\Phi} & \\
    G_{\hat{\theta}}(\hat{A_{\mu}}) & \stackrel{\bar{\nu}}{\rightarrow} & SO(3) &
\end{array}
\label{diagrama2}
\end{eqnarray}
and is also a 2 to 1 homomorphism. If $x \in \mathbb{Z}_{2}^{3}$ then $\psi^{-1}(\lbrace x \rbrace)=\lbrace y_{1},y_{2}\rbrace \subset Q\times \mathbb{Z}_{2}$. Hence:
\begin{eqnarray}
\bar{\nu}(x)&=&\bar{\nu}(\psi(y_{k}))=\bar{\nu} \circ \psi (y_{k})=\Phi \circ \nu (y_{k})\nonumber\\
&=&\Phi(\nu (y_{k}))=\Phi(\nu (q_{k}, g_{k})),
\end{eqnarray}
with $k=1$ or $2$.
Then:
\begin{itemize}
\item $\bar{\nu}$ is an homomorphism.\\
Proof:
\begin{eqnarray}
\bar{\nu}(x'x)&=&\Phi(\nu ((q'_{k}, g'_{k})(q_{l}, g_{l})))=\Phi(\nu (q'_{k}, g'_{k}))\Phi(\nu(q_{l}, g_{l}))\nonumber\\
&=&\Phi \circ \nu (q'_{k}, g'_{k})\Phi \circ \nu(q_{l}, g_{l})=\bar{\nu} \circ\ \psi (q'_{k}, g'_{k})\bar{\nu} \circ \psi(q_{l}, g_{l})\nonumber\\
&=&\bar{\nu}(x')\bar{\nu}(x),
\end{eqnarray}
with $l=1$ or $2$.
\item $\bar{\nu}$ is 2 to 1.\\
Proof: From $\Phi \circ \nu = \bar{\nu} \circ \psi$ and the fact that $\Phi$, $\nu$ and $\psi$ are 2 to 1, it follows that $\bar{\nu}$ is also 2 to 1.
\end{itemize}

Taking into account diagrams (\ref{diagrama1}) and (\ref{diagrama2}), the group homomorphisms:
\begin{equation}
\varphi=\gamma \circ \beta \circ \nu
\end{equation}
and
\begin{equation}
\bar{\varphi}=\bar{\gamma} \circ \bar{\beta} \circ \bar{\nu},
\end{equation}
make commutative the following diagram: 
\begin{eqnarray}
\begin{array}[c]{cccc}   
    G_{\hat{\theta}}(\hat{\psi}) &  \stackrel{\varphi}{\rightarrow} & \bar{\cal{P}}_{4}^{c}\\
    \downarrow\scriptstyle{\psi}&  &\downarrow\scriptstyle{\overset{\approx}{\Phi}} &\\
    G_{\hat{\theta}}(\hat{A_{\mu}})&  \stackrel{\bar{\varphi}}{\rightarrow} & {\cal{P}}_{4}^{c};
\end{array}
\label{diagrama3}
\end{eqnarray}
making explicit the close and possibly deep relationship between these discrete and continuous groups.

\section{Discussion}

\hspace{0.5cm}
In summary, we have that $G_{\hat{\theta}}(\hat{\psi})$ and $G_{\hat{\Theta}}(\hat{A_{\mu}})$, which are groups acting at the quantum field level that include the charge conjugation operator, emerge in a natural way from the $\cal{PT}$-group and its $\cal{P}$ (or $\cal{T}$) subgroups. That is, from matrices acting on Minkowski classical space-time.

It is important to note that $G_{PT}$ generates $G_{\hat{\Theta}}(\hat{A_{\mu}})$, the CPT group of the electromagnetic field, without passing through $SU(2)$. That is, without the need of using spinors; while  the group $SU(2)$ is needed in order to generate $G_{\hat{\theta}}(\hat{\psi})$, the CPT group of the Dirac field.

Finally, another important thing that we found is the relationship between discrete groups, like $G_{\hat{\Theta}}(\hat{A_{\mu}})$ and $G_{\hat{\theta}}(\hat{\psi})$, and continuous groups, like the connected Poincaré group (${\cal{P}}_{4}^{c}$) and its universal covering ($\bar{\cal{P}}_{4}^{c}$). This is shown in diagram (\ref{diagrama3}).

\section{Acknowledgment}
This work was partially support by the project PAPIIT IN 118609-2, DGAPA-UNAM, México. 



\begin{thebibliography}{99}

\bibitem{CPT group}  M. Socolovsky, \emph{ The CPT group of the Dirac Field}, Int J Theor Phys, \textbf{43} (2004), pp. 1941-1967; arXiv: math-ph/0404038. 

\bibitem{B2}  B. Carballo Pérez and M. Socolovsky, \emph{Irreducible representations of the CPT groups in QED}, IJPAM (in press) (2010); arXiv: math-ph/0906.2381v3.

\bibitem{B1}  B. Carballo Pérez and M. Socolovsky, \emph{Charge conjugation from space-time inversion}, Int J Theor Phys, \textbf{48} (2009), pp. 1712-1716; arXiv: hep-th/0811.0842v1.

\bibitem{Azcarraga} J. A. de Azcárraga, \emph{P, C, T, $\theta$ in Quantum Field Theory}, GIFT 7/75 (1975), 69.









\end{thebibliography}
\end{document}